\documentclass[11pt]{article}
\usepackage{moriond,epsfig}

\newcommand{\beq}{\begin{eqnarray}}
\newcommand{\eeq}{\end{eqnarray}}

\newcommand{\centeron}[2]{{\setbox0=\hbox{#1}\setbox1=\hbox{#2}\ifdim
                           \wd1>\wd0\kern.5\wd1\kern-.5\wd0\fi \copy0
                           \kern-.5\wd0\kern-.5\wd1\copy1\ifdim\wd0>\wd1
                           \kern.5\wd0\kern-.5\wd1\fi}}
\newcommand{\ltap}{\>\centeron{\raise.35ex\hbox{$<$}}
                   {\lower.65ex\hbox{$\sim$}}\>}
\newcommand{\gtap}{\>\centeron{\raise.35ex\hbox{$>$}}
                   {\lower.65ex\hbox{$\sim$}}\>}
\newcommand{\gsim}{\mathrel{\gtap}}
\newcommand{\lsim}{\mathrel{\ltap}}

\newcommand\ZZ{\hbox{\zfont Z\kern-.4emZ}}
\font\zfont = cmss10 

\def\ra{\rightarrow}

\def\be{\begin{equation}}
\def\ee{\end{equation}}
\def\bea{\begin{eqnarray}}
\def\eea{\end{eqnarray}}

\begin{document}
\vspace*{4cm}
\title{ELECTROWEAK PRECISION TESTS OF LITTLE HIGGS THEORIES
\footnote{Talk given at XXXVIII Rencontres de Moriond on 
Electroweak Interactions and Unified Theories, Les Arcs 
France, March 15-22, 2003.}}

\author{Graham D. Kribs}

\address{Department of Physics, University of Wisconsin, Madison, WI
53706 USA \\
kribs@physics.wisc.edu}

\maketitle\abstracts{
Little Higgs theories are a fascinating new idea to solve the 
little hierarchy problem by stabilizing the Higgs mass against 
one-loop quadratically divergent radiative corrections.
In this talk I give a brief overview of the idea, focusing
mainly on the littlest Higgs model, and present a sampling of 
the electroweak precision constraints.}

\section{Introduction}

\begin{figure}
\centerline{\includegraphics[width=0.5\hsize]{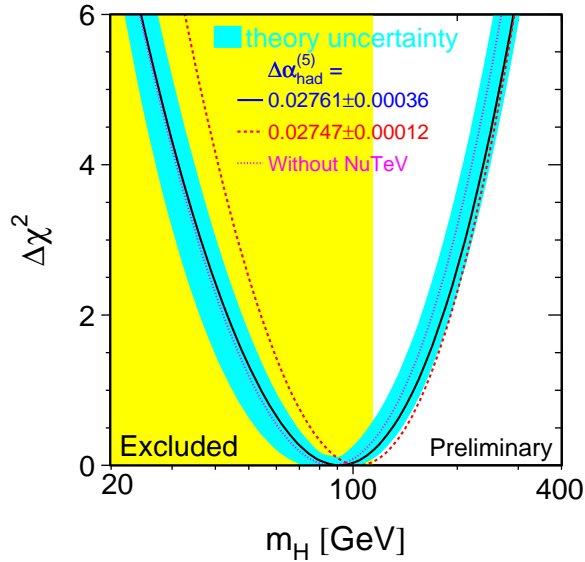}}
\label{higgs-fig} 
\caption{Best fit of the electroweak data as a function of the
Higgs mass.  Taken from Ref.~\protect\cite{LEPEWG}}
\end{figure}

It is well known that the Higgs boson mass is quadratically
sensitive to heavy physics.  The quadratic sensitivity 
arises from Yukawa couplings, gauge couplings, and the Higgs quartic
coupling.  Naturalness suggests the cutoff scale of the Standard Model (SM) 
should be only a loop factor higher than the Higgs mass,
\begin{equation}
\Lambda \lsim 4 \pi m_h \; .
\end{equation}
However, there are many probes of physics beyond the SM at scales ranging 
from a few to tens of TeV\@. In particular, four-fermion operators 
that give rise to new electroweak contributions generally constrain the new
physics scale to be more than a few TeV, and some new flavor-changing
four-fermion operators are constrained even further, to be above
the tens of TeV level.  With mounting evidence \cite{LEPEWG} for the
existence of a light Higgs with mass $\lsim 200$ GeV, 
see Fig.~\ref{higgs-fig},
we are faced with understanding why the Higgs mass is so light 
compared with radiative corrections from cutoff-scale physics that 
appears to have been experimentally forced to be well above the TeV
level.  The simplest solution to this ``little hierarchy problem''
is to fine-tune the bare mass against the radiative
corrections, but this is widely seen as being unnatural.

There has recently been much interest
\cite{little1,minmoose,littlest,gregoirewacker,su6sp6,wacker,schmaltz,%
first,hewett,burdman,wisconsin,dudes,Dib:2003zj,wisc2,%
wackchang,second,nelson}
in a new approach to solving the little hierarchy problem, called
little Higgs models.  These models have a larger gauge group
structure appearing near the TeV scale to which the electroweak
gauge group is embedded.  The novel feature of little Higgs models
is that there are approximate global symmetries that protect the
Higgs mass from acquiring \emph{one-loop} quadratic sensitivity to
the cutoff.  This happens because the approximate global
symmetries ensure that the Higgs can acquire mass only through
``collective breaking'', or multiple interactions.  In the limit
that any single coupling goes to zero, the Higgs becomes an exact
(massless) Goldstone boson. Quadratically divergent contributions
are therefore postponed to two-loop order, thereby relaxing the
tension between a light Higgs mass and a cutoff of order tens of
TeV\@.  Schematically, this can be written as
\begin{equation}
\Lambda \lsim (4 \pi)^2 m_h \; .
\end{equation}
Since I am giving this talk at Moriond, it seems appropriate to borrow a 
French colloquialism by saying that this ``2'' in the exponent is the 
raison d'\^etre of little Higgs theories.

There are now several published little Higgs models;
Table~\ref{models-table} summarizes those that have appeared as of 
this conference.
Little Higgs models can be understood from a variety of perspectives
including deconstruction, analogy to the chiral Lagrangian, etc., 
all of which have appeared in the literature.  In the following I 
have chosen to discuss the operational character of little Higgs models 
by analogy to custodial $SU(2)$ of the SM\@.  This at least provides my 
own (odd?) personal perspective, however imperfect the analogy may be.

The Higgs field of the SM transforms as a complex doublet under $SU(2)_L$,
but more generally as a $\mathbf{4}$ of an $SO(4)$ global symmetry that
rotates the four real scalar fields among themselves.
An $SU(2)$ subgroup of the $SO(4)$ symmetry is gauged, corresponding to 
$SU(2)_L$.  When the Higgs acquires a vacuum expectation value, 
the $SO(4)$ global symmetry is broken to $SO(3) \sim SU(2)$.  This residual, 
or ``custodial'' $SU(2)_c$ is an approximate global symmetry of the 
Higgs sector in the SM.  If $SU(2)_c$ were exact, it would imply an 
interesting relation among the gauge boson masses, namely $M_W/M_Z = 1$.
But, gauging hypercharge is incompatible with custodial $SU(2)$, which 
is manifested at tree-level by the relation
\begin{equation}
\frac{M_W^2}{M_Z^2} = \frac{g^2}{g^2 + g'^2} \; .
\end{equation}
Nevertheless there is a smooth transition in which the global custodial 
$SU(2)$ symmetry is restored when $g' \ra 0$.

Little Higgs theories have several similarities to custodial $SU(2)$
of the SM\@.  In the following, I will use the littlest Higgs 
model \cite{littlest}
as my example.  The littlest Higgs model has a global $SU(5)$ symmetry
with a single scalar non-linear sigma model field in the $\mathbf{15}$
of $SU(5)$ [just as in the SM there is a global $SO(4)$ symmetry with
a single Higgs linearly realized in a $\mathbf{4}$ of $SO(4)$].
The symmetric tensor acquires an expectation value breaking 
global $SU(5)$ down to global $SO(5)$ [just as the Higgs acquiring 
an expectation value breaking global $SO(4)$ down to $SU(2)_c$.]
An $[SU(2)\times U(1)]^2$ subgroup of $SU(5)$ is gauged
[just as an $SU(2)_L$ subgroup of $SO(4)$ was gauged].
The expectation value of the symmetric tensor breaks
$[SU(2)\times U(1)]^2 \ra SU(2)_L \times U(1)_Y$
[just as the Higgs breaks $SU(2)_L \times U(1)_Y \ra U(1)_{em}$].
As the gauge couplings for, say, $SU(2)_1 \times U(1)_1$ are
turned off, $g_1,g_1' \ra 0$, an $SU(3)_1$ subgroup of the $SU(5)$ 
global symmetry is restored [just as turning off hypercharge $g' \ra 0$
restores custodial $SU(2)$].  The same procedure can be done for
$SU(2)_2 \times U(1)_2$.  In either case, the restoration of part of 
the global symmetry by ungauging some part of the gauge symmetry
leads to an ``interesting relation'', for little Higgs this is $m_h \ra 0$
[just as $M_W/M_Z \ra 1$].  Of course the symmetry explanation for the
masslessness of the Higgs is because it represents (four of eight) Goldstone
bosons of spontaneously broken $SU(3)$.  Since one arrives at the 
same result regardless of which set of $SU(2)\times U(1)$ symmetries
are ungauged, one concludes that the Higgs cannot acquire a mass
from interactions only with one gauge coupling.  Hence, the
Higgs does not acquire a quadratic divergence from gauge loops to 
one-loop order because the Higgs is protected by an approximate 
global symmetry.

For there to be no one-loop quadratic divergence from also 
interactions with fermions
(in particular, top quarks) and with itself through the quartic 
coupling, the approximate global symmetries must be respected by
these interactions.  This can be done for the top Yukawa by adding
a vector-like pair of fermions $\tilde{t}, \tilde{t}'$ and arranging
that the top quark acquires a mass through collective breaking.
The details for each model can be found in the literature.
Similarly the quartic coupling must also arise through
collective breaking.  In practice this means there are new
particles near the TeV scale that have the effect of canceling
off the quadratic divergences of the Higgs.  In the littlest Higgs
model, for example, there are four new gauge bosons $W_H^a, B_H$, 
a vector-like pair of quarks that mix to yield the right-handed
top quark, and a new scalar field that is a triplet of $SU(2)_L$.
These states successfully cancel off the quadratic divergences of 
the Higgs so long as their masses are not too far from the TeV scale.

\begin{table}
\begin{center}
\begin{tabular}{l|ccccc}
           & Global &                & \# of light    & Higgs \\
Model Name & Symmetry & Gauge Symmetry & Higgs doublets & triplet vev?
\\ \hline
Minimal Moose \cite{minmoose}  & $SU(3)^8$ & $SU(3)\times SU(2) \times U(1)$ & $2$ & yes \\
Littlest Higgs \cite{littlest} & $SU(5)$ & $[SU(2)\times U(1)]^2$ & $1$ & yes \\
Antisymmetric condensate \cite{su6sp6} & $SU(6)$ & $[SU(2)\times U(1)]^2$ & $2$ & no \\
Simple group \cite{dudes} & $SU(4)^4$ & $SU(4)\times U(1)$ & $2$ & no \\
Custodial SU(2) Moose \cite{wackchang} & $SO(5)^8$ & $SO(5)\times SU(2)\times U(1)$ & $2$ & yes \\
\end{tabular}
\end{center}
\label{models-table}
\caption{Little Higgs models as of March 2003.}
\end{table}

\begin{figure}
\centerline{\includegraphics[width=0.7\hsize]{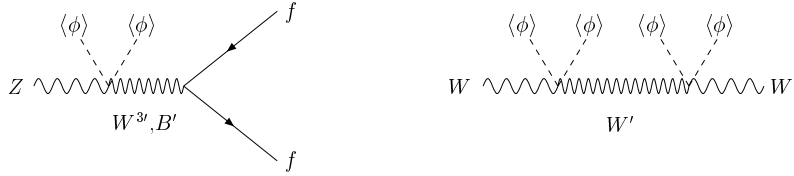}}
\label{naive-fig} \caption{Corrections to $\Gamma_Z$ and to
$M_W^2$ through mixing with heavy gauge bosons.}
\end{figure}

New TeV mass gauge bosons can be problematic if the SM gauge bosons 
mix with them or
if the SM fermions couple to them.  This is because modifications 
of the electroweak sector are usually tightly constrained by 
precision electroweak data (see Refs.~\cite{Sekhar,CST,RSfit,CEKT} 
for example).  Consider
the modification to the coupling of a $Z$ to two fermions and
(separately) the modification to the vacuum polarization of the $W$,
as shown in Fig.~\ref{naive-fig}.  These are among the best
measured electroweak parameters that agree very well with the
SM predictions (using $M_Z$, $G_F$, and $\alpha_{em}$
as inputs):  both of these observables have been measured to
$\pm0.2\%$ to 95\% C.L.  Generically the corrections to these
observables due to heavy $U(1)$ gauge bosons and heavy $SU(2)$ gauge
bosons can be simply read off from Fig.~\ref{naive-fig} as
\begin{equation}
\frac{\delta \Gamma_Z}{\Gamma_Z} \sim 1 + c_1 \frac{v^2}{f^2}
\quad , \quad \frac{\delta M_W^2}{M_W^2} \sim 1 + c_2
\frac{v^2}{f^2} \; ,
\end{equation}
where $f$ is roughly the mass of the heavy gauge boson, $c_1$ and
$c_2$ parameterize the strength of the couplings between
heavy-to-light fields.  For $c_1 \sim 1$ or $c_2 \sim 1$, it is
trivial to calculate the electroweak bound on $f$,
\begin{equation}
f > 5.5 \; \mbox{TeV} \quad \mbox{to 95\% C.L.} \; .
\end{equation}
Notice that even if the coupling of light fermions to the heavy
gauge bosons were zero ($c_1 = 0$), maximal mixing among $SU(2)$
gauge bosons ($c_2 = 1$) is sufficient to place a strong
constraint on the scale of new physics.
In principle one needs only calculate these coefficients 
(and those of other electroweak observables) and combine them
using a global fit to determine the bounds from electroweak 
precision data.  This is straightforward but technical, and 
so I refer interested readers to the original papers \cite{first,second}
for details.

\begin{figure}[t]
\centerline{\includegraphics[width=0.5\hsize]{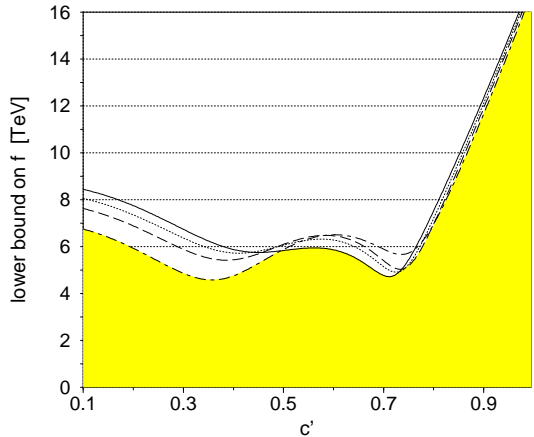}}
\label{limit-fig}
\caption{For the littlest Higgs model with light fermions coupling
only to $SU(2)_1 \times U(1)_1$, the region of parameters excluded to 
95\% C.L. is shown as a function of $c'$.  
The region below the contours is excluded to 95\% C.L. for $c$ equal 
to $0.1$ (solid), $0.5$ (dotted), $0.7$ (dashed), $0.99$ (dot-dashed). 
The shaded region is excluded for any choice of $c$.}
\end{figure}

At this point I should stress that mixing between heavy and light
gauge bosons, as well as the coupling of heavy gauge bosons to
light fermions is a parameter-dependent and/or model-dependent issue 
that does not directly
constrain the little Higgs mechanism for canceling quadratic 
divergences.  Here I shall parameterize simple
examples that will show what direction the electroweak 
constraints tend to ``push'' on the parameter space of models.
Consider the littlest Higgs model with light fermions coupling
to \emph{only} $SU(2)_1 \times U(1)_1$.  The four high energy
gauge groups have four couplings that match onto the well-known
$SU(2)_L \times U(1)_Y$ couplings, leaving two parameters that
we take to be angles defined by $c \equiv \cos\theta = g/g_2$ and 
$c' \equiv \cos\theta' = g'/g'_2$.  
To ensure the high energy gauge couplings $g_{1,2},g'_{1,2}$ are
not strongly coupled, the angles $c,s,c',s'$ cannot be too small.  
We conservatively allow for $c,s,c',s' > 0.1$, or equivalently
$0.1 < c,c' < 0.995$.  We allow the symmetry breaking scale $f$ 
to take on any value (although for small enough $f$ there will be 
constraints from direct production of $B_H$).  The general procedure we 
to systematically step through values of $c$ and $c'$, finding the
lowest value of $f$ that leads to a shift in the $\chi^2$
corresponding to the 68\%, 95\%, and 99\% confidence level (C.L.).
For a three-parameter fit, this corresponds to a $\Delta \chi^2$
of about $3.5$, $7.8$, $11.3$ from the minimum, respectively.
The bound on $f$ is perhaps best illustrated as
a function of $c'$, as shown in Fig.~\ref{limit-fig}.  The shaded
area below the lines shows the region of parameter space excluded by
precision electroweak data.  Note that
we numerically found the value of $c$ that gave the
\emph{least restrictive} bound on $f$ for every $c'$.  For a specific choice
of $c$ the bound on $f$ can be stronger as shown by the different
contours in the figure.

Light fermions that are charged under just $U(1)_1$ maximally couple
to both $B_L$ and $B_H$.  In the littlest Higgs model, $B_H$
is curiously light due to group theoretic factors, and is one
source of the strong constraints on the model.  Although the
top quark must have certain well-defined couplings to $U(1)_1$
and $U(1)_2$ to ensure that the global symmetries are preserved
by its Yukawa interaction, the light fermions have no such 
restriction since their one-loop quadratically divergent contribution 
to the Higgs mass is numerically negligible.  In Ref.~\cite{second}
we considered varying the $U(1)$ charges of the light fermions.
This results in a free parameter
$R$ that characterizes how strongly a given fermion is coupling
to $U(1)_1$ ($R$ times its hypercharge) and $U(1)_2$ ($1-R$ times
its hypercharge).  If we wish to maintain integer powers of the 
non-linear sigma model field when writing Yukawa couplings, then $R$ can
only take on fractional integer powers $0, 1/5, 2/5, \ldots$.
Three interesting cases are $R=1$, the choice in Fig.~\ref{limit-fig};
$R=4/5$, the choice that leads to dimension-4 Yukawa couplings;
and $R=3/5$, the choice that is identical to that of the top quark.

\begin{figure}[t]
\centerline{\includegraphics[width=1.0\hsize]{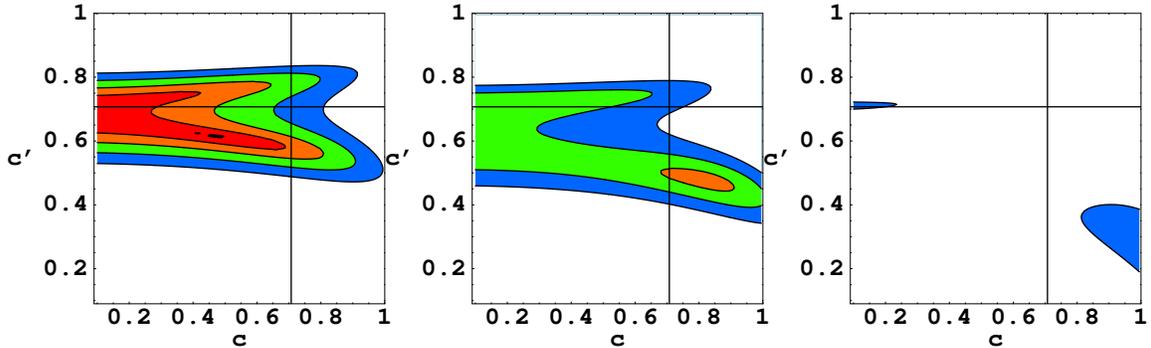}}
\caption{Contours of the minimum allowed values of $f$ at 95\% C.L.
for $a=1$ and $R=3/5$ (left graph) $R=4/5$ (center graph) $R=1$ (right graph).
The black contour is for $f<1$ TeV, red is $f<2$ TeV, orange is $f<3$ TeV,
green is $f<4$ TeV, blue is $f<5$ TeV, and white is $f>5$ TeV\@.}
\label{fifths}
\end{figure}

In Fig.~\ref{fifths} a contour plot for fixed $R=3/5,4/5,1$ shows 
the allowed range
of parameter space at 95\% C.L. for both $c$ and $c'$ showing the
size of the allowed region of parameter space for a given value of
$R$.  Unlike what we found for $R=1$, it is clear that for $R=3/5$
there are restricted regions of parameter space where the bound on $f$
is in the $1$-$2$ TeV.
This illustrates that varying the strength of the coupling of light 
fermions to $B_H$ leads to quite dramatically different constraints on 
the littlest Higgs model.  

\begin{figure}[t]
\centerline{\includegraphics[width=0.35\hsize]{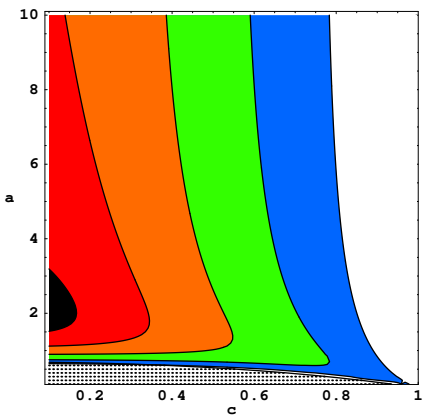}}
\caption{Contour plot of the allowed values of $f$ as a function
of the parameters $a$ and $c$.  The
allowed contours correspond to less than $1,2,3,4,5$ TeV\@.  The
grey shaded region at the bottom corresponds to the region
excluded by requiring a positive triplet (mass)$^2$.} 
\label{oneu1a}
\end{figure}

One alternative that avoids all of the difficulties associated
with $B_H$ is to simply gauge $U(1)_Y$.  This leads to a one-loop
quadratic divergence proportional to $g'^2$, but this is numerically
small if the cutoff scale is around 10 TeV\@.  Now it becomes more
important to examine a fuller set of contributions to the electroweak 
observables.  There are (at least) two additional effects:
in the littlest Higgs model there is an $SU(2)_L$ triplet scalar
field that acquires a vev $v' = {\cal O}(v^2/f)$, leading to 
an additional tree-level contribution, while at one-loop there
are significant contributions from the heavy top.  For this 
discussion let me consider only the triplet vev in addition to
heavy gauge boson exchange.  The triplet vev can be calculated
by minimizing the effective potential; the details are in
Refs.~\cite{littlest,first}.  One finds 
\begin{equation}
|v'|^2 = \frac{v^4}{16 f^2} \left[
\frac{2 \lambda}{a(g_1^2+g_1'^2)}-1\right]^2~. \label{deltaprime}
\end{equation}
in terms of a single parameter $a \sim {\cal O}(1)$ in the 
Coleman-Weinberg effective 
potential and the Higgs quartic coupling $\lambda$.  We can then 
recalculate the electroweak observables including a triplet vev, 
and for a littlest-type Higgs model in which only $SU(2)_1 \times 
SU(2)_2 \times U(1)_Y$ is gauged the result shown in
Fig.~\ref{oneu1a}.
Interestingly, for suitable $SU(2)_{1,2}$ gauge couplings
$g_2 \gg g_1$ and an order one parameter in the Coleman-Weinberg 
potential, the bound on $f$ is $1$-$2$ TeV.

There are several other variations of the littlest Higgs model,
and variations in the global symmetries that lead to other
interesting constraints.  Here I restricted myself to presenting
only a small sampling of the little Higgs models or variations, and their
electroweak constraints.  There are a few general lessons that are
already clear:  First, for generic choices of couplings of light fermions 
to heavy gauge bosons (such as coupling to just $SU(2)_1 \times U(1)_1$), 
or generic light/heavy gauge boson mixing (such as $c \sim 1/\sqrt{2}$),
the bound on the symmetry breaking scale $f$ is maximized to of order 
$4$-$5$ TeV\@.  In the littlest Higgs model, a lower bound on 
$f$ can be translated into a lower bound on the mass the vector-like 
quarks $m \gsim \sqrt{2} f$, and then translated into the minimal amount 
of fine-tuning to obtain a light Higgs.  However, decoupling
the light fermions from the heavy $B$ gauge boson (either exactly
or approximately as shown above for $R=3/5$) or eliminating the
heavy $B$ gauge boson entirely from the spectrum, while choosing the
observed $W$ gauge boson to be nearly pure $W_1$ generally allows the 
symmetry breaking 
scale $f$ to be $1$-$2$ TeV\@.  In some models, bounding $f$ does 
not directly imply an upper bound on the amount of fine-tuning;
other parameters in the theory that can be suitably chosen to,
for example, keep 
the heavy vector-like quarks significantly lighter than their heavier 
gauge boson cousins.  We can surely expect exciting further 
developments in new little Higgs models and UV completions!

\section*{Acknowledgments}

It is a pleasure to thank Csaba Cs\'aki, Jay Hubisz, 
Patrick Meade, and John Terning for collaboration leading to the
papers \cite{first,second} on which this talk was based.
I also thank the organizers of the XXXVIII Rencontres de Moriond 
(Electroweak Session) 
for the invitation to participate and by providing partial financial 
support through an NSF grant to travel to the conference.  Finally,
my research was supported by the US Department of Energy under 
contract DE-FG02-95ER40896.

\end{document}